\documentclass[twocolumn,aps,prb,floatfix,showpacs]{revtex4}

\usepackage{graphics}

\begin{document}

\title{Appearance of Universal Metallic Dispersion in a Doped Mott
Insulator}

\author{S. Sahrakorpi$^{1}$, R.S. Markiewicz$^{1}$, Hsin Lin$^{1}$,
M. Lindroos$^{1,2}$, X. J. Zhou$^{3,4}$, T. Yoshida$^{5}$,
W. L. Yang$^{3,4}$, T. Kakeshita$^{6}$, H.  Eisaki$^{3,6}$,
S. Uchida$^{6}$, Seiki Komiya$^{7}$, Yoichi Ando$^{8}$, F. Zhou$^{9}$,
Z. X. Zhao$^{9}$, T. Sasagawa$^{3,10}$, A.  Fujimori$^{5}$,
Z. Hussain$^{4}$, Z.-X. Shen$^{3,4}$, and A. Bansil$^1$}

\affiliation{$^1$Physics Department, Northeastern University, Boston,
Massachusetts 02115}

\affiliation{$^2$Institute of Physics, Tampere University of
Technology, P.O. Box 692, 33101 Tampere, Finland}

\affiliation{$^{3}$Dept. of Physics, Applied Physics and Stanford
Synchrotron Radiation Laboratory, Stanford University, Stanford, CA
94305}

\affiliation{$^{4}$Advanced Light Source, Lawrence Berkeley National
Lab, Berkeley, CA 94720}

\affiliation{$^{5}$Dept. of Physics, University of Tokyo, Bunkyo$-$ku,
Tokyo 113-0033, Japan
}

\affiliation{$^{6}$Dept. of Superconductivity, University of Tokyo,
Bunkyo$-$ku, Tokyo 113, Japan}

\affiliation{$^{7}$Central Res. Inst. of Electric Power Ind., 2-11-1
Iwato-kita, Komae, Tokyo 201-8511, Japan}

\affiliation{$^{8}$Institute of Scientific and Industrial Research,
Osaka University, Ibaraki, Osaka 567-0047, Japan}

\affiliation{$^{9}$National Lab for Superconductivity, Inst. of
Physics, Chinese Acad. of Sci., Beijing 100080, China}

\affiliation{$^{10}$MSL, Tokyo Institute of Technology,
Kanagawa 226-8503, Japan}

\date{\today}
\begin{abstract}

We have investigated the dispersion renormalization $Z_{disp}$ in 
La$_{2-x}$Sr$_x$CuO$_4$ (LSCO) over the wide doping range of 
$x=0.03-0.30$, for binding energies extending to several hundred meV's.  
Strong correlation effects conspire in such a way that the system exhibits 
an LDA-like dispersion which essentially `undresses' ($Z_{disp}\rightarrow 
1$) as the Mott insulator is approached. Our finding that the Mott 
insulator contains `nascent' or `preformed' metallic states with a 
vanishing spectral weight offers a challenge to existing theoretical 
scenarios for cuprates.

\end{abstract}


\maketitle

Under strong electronic correlations the parent compounds of all
cuprate high-temperature superconductors assume the so-called
Mott-Hubbard insulating state, rather than the conventional metallic
state. By what routes these insulators accomplish the miraculous
transformation into superconductors with the addition of electrons or
holes is a question of intense current interest, which bears on
ongoing debates surrounding the interplay between electron
correlations, magnetism, lattice effects, and the mechanism of
high-temperature superconductivity.\cite{Damascelli2003} In this study
we consider the classic superconductor La$_{2-x}$Sr$_x$CuO$_4$ (LSCO)
over the wide doping range of $x=0.03-0.30$, delineating how the
electronic spectrum evolves with doping for binding energies extending
to several hundred meV's. Our analysis indicates that this Mott
insulator contains `nascent' or `preformed' metallic states, which
develop finite spectral weight with doping, but otherwise undergo
relatively little change in dispersion over a wide doping range. Our
findings challenge existing theoretical scenarios for cuprates.

We have carried out extensive angle-resolved photoemission (ARPES)
measurements from LSCO single crystals covering a wide range of dopings,
momenta and binding energies. Although the incoherent part of the spectrum
behaves quite anomalously, we find that many-body effects conspire in such
a way that insofar as the coherent part of the spectrum is concerned, at
least its underlying dispersion is reasonably described by the
conventional band-theory picture, significantly broadened lineshapes and
`kinks' in the dispersion notwithstanding. Surprisingly, even with the
addition of just a few percent holes in the insulator, the full-blown
metallic spectrum seemingly turns on with little renormalization of the
dispersion.  In particular, the spectrum displays the presence of the
tell-tale van Hove singularity (VHS) whose location in energy and
three-dimensionality are in accord with the band theory predictions.
Furthermore, this metallic spectrum is `universal' in the sense that it
depends weakly on doping, in sharp contrast to the common expectation that
dispersion is renormalized to zero at half-filling.

The band structure results  
are based on all electron, full-potential computations 
within local density approximation (LDA) 
using the tetragonal lattice
structure\cite{Jorgensen1987}, and include effects of La/Sr
substitution within the framework of the virtual crystal
approximation\cite{Lin2006}. The ARPES
measurements were carried out on Beamline 10.0.1 at the ALS using
Scienta~200,~2002, and~R4000 electron energy analysers for 55~eV light
with strong in-plane polarization. The energy resolution is 15-20~meV,
and the angular resolution is 0.3~degrees for the 14~degrees angular
mode. All data were taken at T=20K.

\begin{figure}
\begin{center}
\resizebox{8.5cm}{!}{\includegraphics{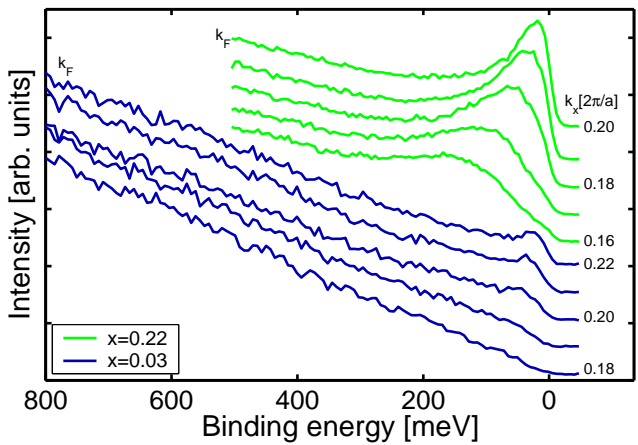}}
\end{center}
\caption{
(Color online) Illustrative ARPES spectra as a function of binding energy in
La$_{2-x}$Sr$_x$CuO$_4$ for a series of momenta along the nodal (i.e.
$\Gamma$ to $(\pi,\pi)$) direction. Results from a lightly doped
insulating sample ($x=0.03$, dark blue lines) and an overdoped metallic
sample ($x=0.22$, light green lines) are shown. Coherent spectral peak is seen
to disperse to higher binding energies as one moves away from the
Fermi momentum, $k_F$.
\label{VHSfig:1}}
\end{figure}

We set the stage for our discussion with the help of Fig.~\ref{VHSfig:1}, which
shows typical spectra from LSCO in the form of energy distribution
curves (EDCs) at two different dopings for a series of
momenta. Considering the overdoped case (upper light green set), we see a
coherent feature dispersing to higher binding energies and becoming
broader as one moves away from the Fermi momentum $k_F$. This feature
sits on top of a substantial incoherent background extending to quite
high energies at all momenta. These basic characteristics are seen to
persist in the lightly doped sample (lower dark blue set), although the
greatly reduced spectral weight of the coherent feature in relation to
the incoherent part of the spectrum is very evident. Our focus in this
article is on the aforementioned coherent feature in the spectrum of
LSCO, and especially on delineating the evolution of its dispersion
with doping.

The existence of a large, Luttinger-like, metallic `nascent' or `underlying' Fermi
surface in LSCO has been established in previous studies, culminating
in the recent systematic analysis of Sahrakorpi {\it et al.}\cite{Sahrakorpi2005}
and Yoshida {\it et al.}\cite{Yoshida2006}.  In contrast, here we consider spectra
over a wide energy range of several hundred meV's, show the presence
of the VHS--a unique feature of the band structure--even in the
lightly doped insulator, and establish unequivocally the existence of
near-universal metallic dispersion in LSCO. These `nascent' Fermi
surfaces and dispersions are well defined despite the difficulties of
identifying features in the face of loss of spectral weight as the
pseudogap develops with underdoping. We emphasize that our focus is on
what we may call the `gross' spectrum. In other words, we are not
concerned with the fine structure in the electronic spectrum
associated with the well-known low energy
kinks\cite{Lanzara2001}, the recently discovered
features at higher energy scales,\cite{Ronning2005,Graf2007,Meevasana2007,Xie2007,Valla2007} or
superconducting\cite{Sensarma2007}, or other\cite{Kanigel2006} leading-edge
gaps.

\begin{figure*}[th]
\begin{center}
\resizebox{16.5cm}{!}{\includegraphics{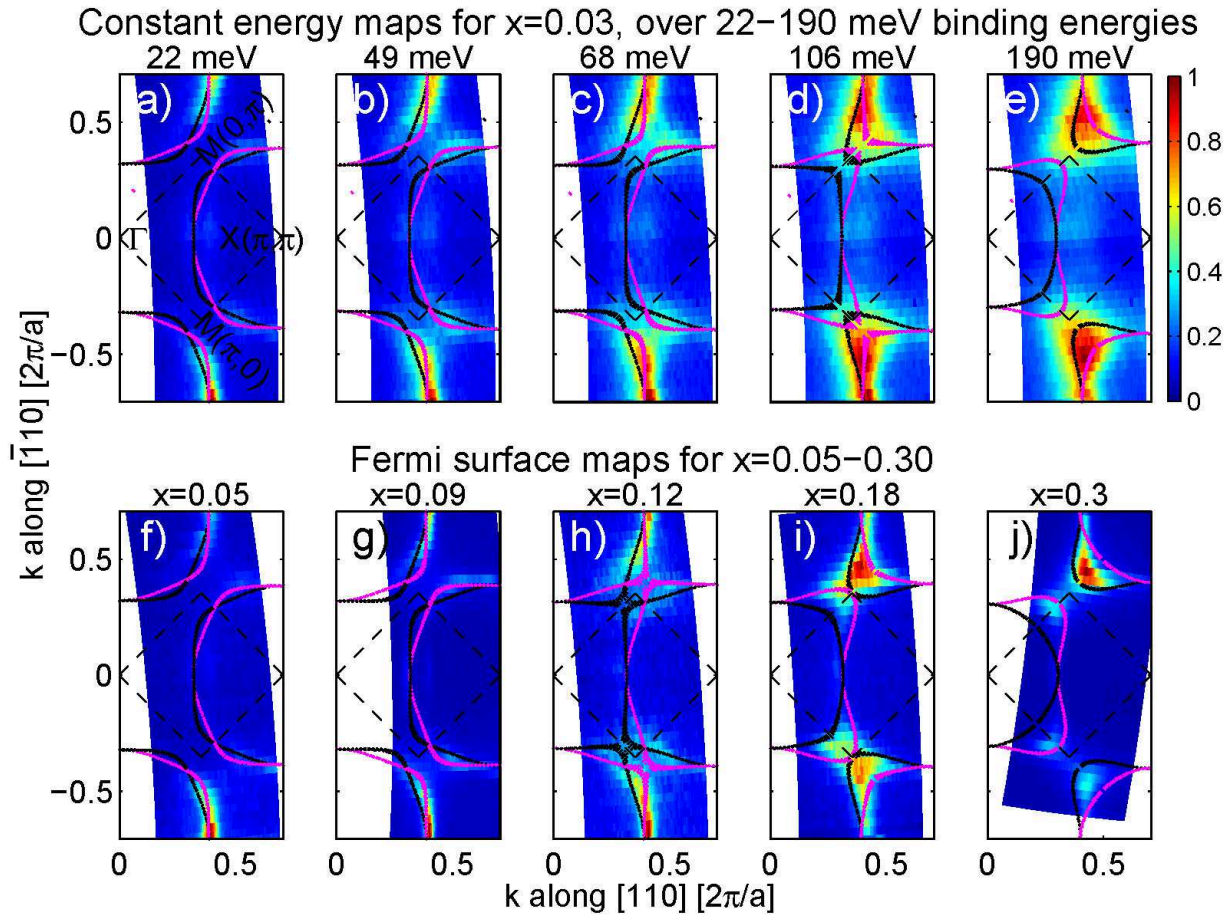}}
\end{center}
\caption{
(Color) Experimental ARPES intensity maps in LSCO are compared with the
corresponding cuts in the $(k_x,k_y)$ plane through the theoretical
constant energy (CE) surfaces for $k_z=0$ (magenta lines), and
$k_z=2\pi/c$ (black lines). {\em Top Row:} $x=0.03$ with binding
energy varying from 22 meV to 190 meV. {\em Bottom Row:} ARPES maps
for emission from the Fermi energy for dopings varying from $x=0.05$
to $x=0.30$.
\label{VHSfig:2}}
\end{figure*}

The top row of Fig.~\ref{VHSfig:2} shows ARPES intensity maps from a lightly doped
sample of LSCO ($x=0.03$) for a series of binding
energies. Cross-sections of the corresponding constant energy (CE)
surfaces in the $(k_x,k_y)$ plane computed from the band structure of
LSCO, superposed at $k_z=0$ (magenta lines) and $k_z=2\pi/c$ (black
lines) indicate the expected broadening of the ARPES spectra
associated with interlayer coupling. At zero binding energy, i.e. the
Fermi energy $E_F$, such CE contours give the projection of the 3D FS
of LSCO on to the $(k_x,k_y)$ plane.  Notably, the momentum region
enclosed by these CE contours defines the region of allowed ARPES
transitions, modulated by the effect of the ARPES matrix
element.\cite{Bansil1999matrix,Bansil2005,Sahrakorpi2005} At low binding energies, the
CE surface is seen from panels (a-c) to be hole-like around the
$X(\pi,\pi)$ point for all $k_z$ values. In contrast, at high energy
in panel (e), after the VHS has been crossed, the CE surface becomes
completely electron-like centered at $\Gamma$. The transition from a
hole- to electron-like CE surface does not take place abruptly because
the VHS possess a significant 3D character, extending from 85-140~meV
in binding energy.

The evolution of the experimental ARPES intensity pattern with binding
energy in the top row of Fig.~\ref{VHSfig:2} clearly follows that of the projected
CE surfaces. In particular, the spectral intensity remains confined
mainly within the boundaries of these projections as expected, and
with increasing binding energy, the intensity first spreads towards
the $M(\pi,0)$-points and then moves away from the $M$-points along a
perpendicular direction, very much the way the CE surfaces transition
from being hole- to electron-like. Moreover, 
first principles ARPES computations show that under the combined
effects of the matrix element and {$k_z$} dispersion, the spectral
intensity develops the characteristic 'wing-like' shape seen in
Fig.~\ref{VHSfig:2}, and that the spectral weight grows rapidly in the antinodal
region as the VHS is approached.\cite{Sahrakorpi2005} These results
leave no doubt that metallic states, including the presence of the 3D
VHS, appear in the spectrum of the insulator with the addition of only
a few percent holes, even though spectral broadening and incoherent
background make it hard to see this directly in the individual
EDCs. The observed location in energy and three-dimensionality of the
VHS is well-described by the conventional band theory picture,
indicating that the energies of these metallic states undergo
little renormalization in the lightly doped insulator. 
The evolution of the VHS over a
denser binding energy mesh can be seen further in the three movies for
$x=0.03, 0.07,$ and $0.12$ provided as supplementary material.\cite{footnote}
The characteristic dispersion of the VHS is quite recognizable in the 
emission maps of the top row of Fig.~\ref{VHSfig:2}.
and in the supplementary movies.

\begin{figure*}[th]
\begin{center}
        \resizebox{16.5cm}{!}{\includegraphics{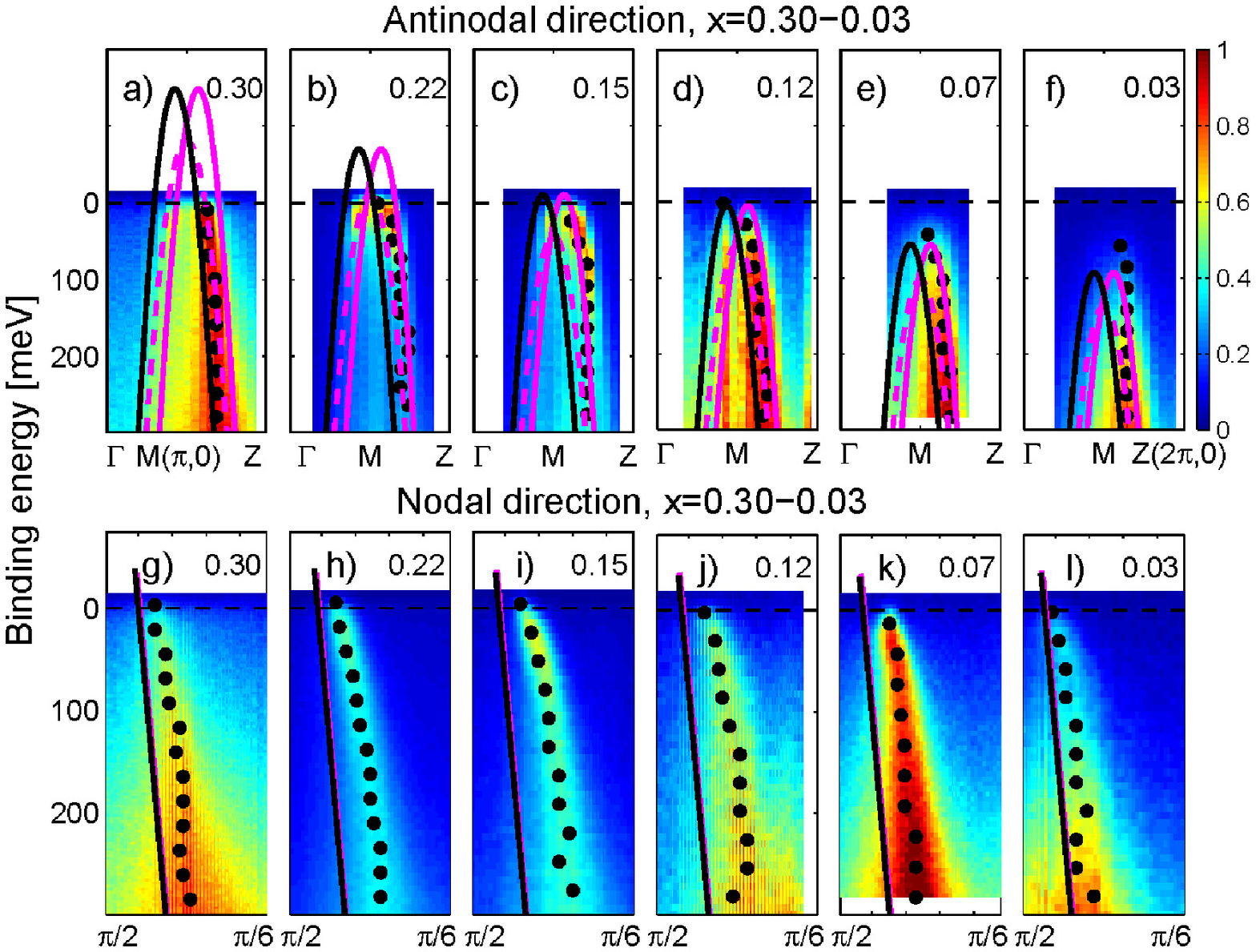}}
\end{center}
\caption{
(Color) ARPES intensity maps along the antinodal (top row) and nodal (bottom
row) directions over the doping range $x=0.30-0.03$. The corresponding
computed band structures are also plotted for three different values
of $k_z$: $k_z=0$ (magenta solid line), $k_z=\pi/c$ (magenta dashed
line), and $k_z=2\pi/c$ (black solid line). Black dots mark positions
of the peaks in the experimental spectra. Note that nodal data in the
lower
row are plotted on an
expanded horizontal scale in order to highlight relatively small
differences between gross theoretical and experimental dispersions. For 
each
doping,
the nodal and antinodal intensities are normalized to a common maximum.
\label{VHSfig:3}}
\end{figure*}

We demonstrate next that the aforementioned metallic dispersion is
only weakly dependent on doping. Note that if this is true then the
main difference in going from one doping to another would be a shift
in the Fermi energy needed to accommodate the right number of holes in
the filled portion of the band structure. That is, topologies of the
CE surfaces and the associated emission spectra would be comparable
for various dopings except for a rigid shift of the energy
scales. That this is indeed the case is shown by the results for
emission from the $E_F$ for $x=0.05-0.30$, presented in the bottom row
of Fig.~\ref{VHSfig:2}. For example, based on parameter free LDA computations for
$x=0.03$ and $x=0.30$, the $E_F$ for $x=0.30$ is lower by 190~meV than
for $x=0.03$, so that the ARPES map for emission at a binding energy
of 190~meV from $x=0.03$ in panel~(e) can be compared with that for
emission from the $E_F$ for $x=0.30$ in~(j). In this vein, the binding
energies in the various panels of the top row of Fig.~\ref{VHSfig:2} for $x=0.03$
have been chosen to match the $E_F$ shifts involved at the doping
levels considered in the panels of the bottom row. Good accord is seen
in all cases.  
We emphasize that the energy shifts corresponding to the respective panels of the top and bottom rows in Fig.~\ref{VHSfig:2} are based on parameter free first principles computations. 
These and other similar comparisons among spectra taken
at different binding energies and doping levels show clearly that LSCO
is characterized by a near-universal metallic dispersion
despite dramatic changes in the lineshape
due to interactions. This is also true for the theoretical
dispersions, although slight doping dependencies can be seen for
example by comparing the CE surfaces in the top and bottom frames~(c)
and~(h) in Fig.~\ref{VHSfig:2}.

Further insight is provided by Fig.~\ref{VHSfig:3}, which shows plots of spectral
intensity as a function of binding energy along the antinodal line
(top row) and the nodal line (bottom row) for six different
dopings. The corresponding energy bands at $k_z=0$, $\pi/c$ and
$2\pi/c$ are overlaid in each panel. These bands are seen to differ
substantially for different $k_z$ values along the antinodal line in
the upper panels but are virtually indistinguishable in the lower
panels along the nodal line. The VHS, which is marked by the extremum
of the band along the antinodal line, is spread over 40~meV at
$x=0.03$ doping due to the effect of $k_z$ dispersion, and its center
lies at 110~meV below $E_F$ for $x=0.03$, but moves to 110~meV above
the $E_F$ for $x=0.30$ with a spread of 70~meV.  Along the nodal
direction, the theoretical band follows the experimental peak
positions given by the black dots reasonably well, although the
experimental points are shifted to the right compared to theory in
most cases, indicating a slight deviation of FS shape from LDA.
Along the antinodal direction, however, the spectral peaks
are substantially broader due to $k_z$ dispersion and also possible
many body interactions, although most of the spectral weight lies
within the k$_z$ dispersed bands as expected in a quasi-2D
system.\cite{Bansil1999matrix,Bansil2005,Sahrakorpi2005}

\begin{table}
\begin{center}
\begin{tabular}{|l|c|c|c|c|c|c|}
\hline
Doping $x=$	&0.03&	0.07&	0.12&	0.15&	0.22&	0.30 \\ \hline
$Z_{disp}$ (nodal)&	1.0	& 0.8 	&0.6 	& 0.6 	& 0.6 &	0.6 \\ 
&	 (1.2)	& (1.2)	& (0.8)	& (0.8)	&  (0.7)&	 (0.6) \\ \hline
$Z_{disp}$ (antinodal)&	1.1$\pm$0.2	&1.3$\pm$0.4	&$>$0.6$^*$ &	$>$0.6$^*$ &	$>$0.7$^*$ &	$>$0.7$^*$  \\ \hline
\end{tabular}
\end{center}
\caption{Estimates of dispersion renormalization factors $Z_{disp}$ in LSCO for different dopings $x$ in relation to the LDA values. Nodal $Z_{disp}$ values in parenthesis are an estimate of the upper limit. Stars in the second row denote that for these dopings the values refer to the region in the vicinity of the antinodal point as discussed in the text.
\label{tab:Zdisp}
}
\end{table}

We have examined the renormalization of dispersion in relation to the
LDA values along the nodal and antinodal directions as a function of
doping, and thus obtained the associated renormalization factors
$Z_{disp}$. These results are summarized in Table~\ref{tab:Zdisp}. 
Our study provides new insight into the value of $Z_{disp}$ along the antinodal direction because the VHS is a very robust feature of the LDA band structure. By determining the position of the VHS in the experimental spectrum, and comparing this position with that expected from the LDA, we can uniquely determine the overall renormalization of the spectrum along the antinodal direction with respect to the LDA. At $x=0.03$ and $x=0.07$, the VHS lies below the Fermi energy for all $k_z$ values, and in these two cases, by analyzing the supplementary movies\cite{footnote}
using different renormalization factors to scale the LDA bands, we have obtained $Z_{disp}$ (antinodal) values of $\sim 1$ within the uncertainty shown in Table~\ref{tab:Zdisp}, so that the LDA bands are essentially unrenormalized. For $x=0.12$ and higher dopings, part or all of the VHS lies above the Fermi energy, so that we were only able to estimate a lower limit for $Z_{disp}$ in the vicinity of the antinodal point. There are of course no filled states at the antinodal point once the VHS moves above the Fermi energy.

\begin{figure}
\begin{center}
        \resizebox{8.5cm}{!}{\includegraphics{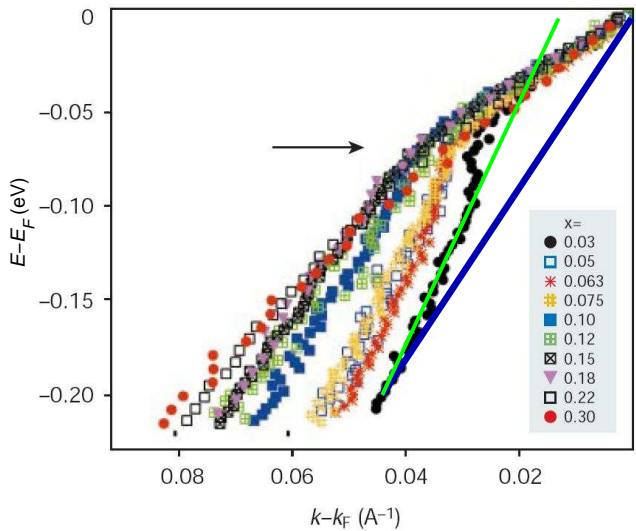}}
\end{center}
\caption{
(Color online) Reproduction of Fig.~1 in Zhou {\it et al.}\cite{Zhou2003} Dark blue line drawn on the $x=0.03$ dataset illustrates how the nodal $Z_{disp}$ value is defined here, while the light green line gives the high energy slope used to estimate the corresponding upper limit on $Z_{disp}$ (nodal) as discussed in the text. 
\label{VHSfig:S1}}
\end{figure}

Turning to the nodal direction, we note that here the effects of $k_z$ dispersion are small, and our analysis is consistent with results available in the literature. For completeness, however, we have estimated the gross $Z_{disp}$ (nodal) values from the nodal spectra given in Fig.~1 of Zhou {\it et al.}\cite{Zhou2003}, reproduced here as Fig.~\ref{VHSfig:S1}. 
In order to gain a handle on the `gross' or underlying dispersion exclusive of the low energy kink 
(see arrow at 70~meV in Fig.~\ref{VHSfig:S1}), values of
$Z_{disp}$ (nodal) given in Table~\ref{tab:Zdisp} are
obtained from the slopes of the straight lines joining the point at
the Fermi energy with that at 200 meV binding energy, i.e. by lines
such as the dark blue line drawn for the $x=0.03$ case in
Fig.~\ref{VHSfig:S1}, and comparing this slope to the corresponding
LDA values. Since the LDA values of the nodal radii do not exactly
match the measured values, we have estimated $Z_{disp}$ (nodal) by
comparing LDA and experimental slopes.
We also show in parentheses the values of $Z_{disp}$ (nodal) obtained from the slopes of the high energy part of the spectrum, as given for example by the light green line for $x = 0.03$ in Fig.~\ref{VHSfig:S1}, which provides an estimate for the upper limit of $Z_{disp}$ (nodal).  

Interestingly, at the lowest doping of $x = 0.03$, the nodal as
well as the antinodal renormalization factor is seen to be $\sim$1. In
the optimally and overdoped regimes, the nodal renormalization factor
is $\sim$0.6, while the value of the antinodal factor is estimated to
be greater than 0.6. In the underdoped $x = 0.03$ case, the
renormalization of states in the antinodal and nodal directions is
roughly similar, but this is less clear at higher dopings.  These
results are surprising since we might have expected the LDA to
provide a reasonable description in the overdoped regime, and to be
renormalized greatly in the underdoped case.

Even though we have shown that the gross dispersion up to several
hundred meV's is described quite well by the band theory picture, we
emphasize that this does not mean that the spectrum of LSCO is
conventional in nature. 
As shown in Fig.~\ref{VHSfig:1}, the coherent spectral weight of these
dispersive features fades away for underdoping where
the pseudogap and polaronic effects kick in.
The spectral weight is
transferred to an incoherent feature\cite{Yoshida2003,Shen2004,Rosch2005},
reflecting the importance of many-body physics.
Our analysis thus
suggests that the spectrum of the insulator already contains
`preformed' or `nascent' metallic states, which possess zero spectral
weight in the half-filled case. With doping, these states develop
finite spectral weight, but otherwise undergo relatively little change
in their basic character.

It is interesting to study the relationship between the renormalization 
of the dispersion and spectral weight within the simple Green function formalism.
The Green function can be approximately written as 
$G = Z_{\omega} / [\omega -Z_{\omega}\epsilon_k / Z_k]$, with  
$Z_{\omega} = 1 / (1 - \partial\Sigma /\partial\omega )$, and 
$Z_k=1/(1 + \partial\Sigma / \partial\epsilon_k )$, where $\Sigma$ is the self-energy. 
While the spectral weight renormalization is associated with $Z_{\omega}$, 
the band dispersion renormalization is given by 
$Z_{disp}=Z_{\omega} / Z_k$.
It is striking that the spectral weight in LSCO is renormalized
greatly but the dispersion is not. This is not consistent with a
momentum independent self-energy, since the renormalization of
spectral weight would then be accompanied by a similar renormalization
of dispersion, suggesting that the self-energy possesses momentum as
well as energy dependencies such that the effects of the associated
renormalization factors, $Z_\omega$ and $Z_k$,\cite{Paramekanti2004}
conspire to approximately cancel each other in renormalizing
dispersion. 
This picture however implies that the heat capacity would
be renormalized only weakly with doping, which is not the case
experimentally.\cite{Matsuzaki2004}

Insofar as theoretical models of strongly correlated systems are
concerned, a focus of attention in the literature has been the $t-J$
model, which has been used to describe the physics of the Hubbard
Hamiltonian in the large $U$ regime. One class of $t-J$ models finds
gapped insulating solutions at half-filling, which evolve with doping
to yield small Fermi surface pockets\cite{Prelovsek2002}. This route appears
to be followed in $Nd_{2-x}Ce_xCuO_4$ for electron
doping\cite{Armitage2001,Kusko2002,Tremblay2006}. 
But the present results do not show clear evidence for small pockets
or other strong violations of Luttinger's theorem\cite{Kokalj2007}.
On the other hand, metallic
solutions with a large Luttinger-like Fermi surface have also been
reported \cite{Edegger2006,Sensarma2007}, where the spectral weight $Z_\omega$ is
renormalized to zero at half-filling. These solutions are closer to
our observations, even though the band width is found to be strongly
renormalized near half-filling, $Z_{disp}\sim J/t$.  A recent variant of the $t-J$ model
succeeds in capturing both the metallic and insulating aspects of the
spectrum simultaneously\cite{Ribeiro2005,Tan2007}. However, to our knowledge,
our finding of a large Fermi surface and a large dispersion, which
is weakly dependent on doping, does not fit well within the framework
of the currently
available scenarios based on simple Hubbard or $t-J$ models or naive
application of the classical polaron theory.

In conclusion, we have shown that metallic states appear in the
electronic spectrum of high temperature superconductor
La$_{2-x}$Sr$_x$CuO$_4$ at all doping levels, ranging from the lightly
doped insulating to the overdoped metallic regime. The gross
dispersion of these states is not only near-universal in that it is weakly
dependent on doping, but it is also conventional in that it is reasonably
described by the standard band theory picture up to binding energies
of several hundred meV's. However, the metallic states suffer substantial
loss
of spectral weight with underdoping and in this respect behave quite
unconventionally. Our findings challenge existing theoretical models
of the cuprates and indicate the complexity with which many-body
physics plays out in this fascinating material.

{\bf Acknowledgments}
This work was supported by the US Department of Energy (DOE),  
Office of Basic Energy Sciences, with contracts 
DE-FG02-07ER46352 and DE-AC03-76SF00098, and benefited from the
allocation of supercomputer time at NERSC and Northeastern
University's Advanced Scientific Computation Center (ASCC).  The
Stanford work was supported by DOE Office of Science, Division of
Materials Science, with contract DE-FG03-01ER45929-A001. 
A.F. and S.U. acknowledge a Grant-in-Aid for Scientific
Research in Priority Area ``Invention of Anomalous Quantum Materials''
from MEXT, Japan for financial support.
Y.A. was supported by KAKENHI 16340112 and 19674002.


\begin{thebibliography}{29}
\expandafter\ifx\csname natexlab\endcsname\relax\def\natexlab#1{#1}\fi
\expandafter\ifx\csname bibnamefont\endcsname\relax
  \def\bibnamefont#1{#1}\fi
\expandafter\ifx\csname bibfnamefont\endcsname\relax
  \def\bibfnamefont#1{#1}\fi
\expandafter\ifx\csname citenamefont\endcsname\relax
  \def\citenamefont#1{#1}\fi
\expandafter\ifx\csname url\endcsname\relax
  \def\url#1{\texttt{#1}}\fi
\expandafter\ifx\csname urlprefix\endcsname\relax\def\urlprefix{URL }\fi
\providecommand{\bibinfo}[2]{#2}
\providecommand{\eprint}[2][]{\url{#2}}

\bibitem[{\citenamefont{Damascelli et~al.}(2003)\citenamefont{Damascelli,
  Hussain, and Shen}}]{Damascelli2003}
\bibinfo{author}{\bibfnamefont{A.}~\bibnamefont{Damascelli}},
  \bibinfo{author}{\bibfnamefont{Z.}~\bibnamefont{Hussain}}, \bibnamefont{and}
  \bibinfo{author}{\bibfnamefont{Z.~X.} \bibnamefont{Shen}},
  \bibinfo{journal}{Reviews of Modern Physics} \textbf{\bibinfo{volume}{75}},
  \bibinfo{pages}{473} (\bibinfo{year}{2003}).

\bibitem[{\citenamefont{Jorgensen et~al.}(1987)\citenamefont{Jorgensen,
  Schuttler, Hinks, Capone, Zhang, Brodsky, and Scalapino}}]{Jorgensen1987}
\bibinfo{author}{\bibfnamefont{J.~D.} \bibnamefont{Jorgensen}},
  \bibinfo{author}{\bibfnamefont{H.~B.} \bibnamefont{Schuttler}},
  \bibinfo{author}{\bibfnamefont{D.~G.} \bibnamefont{Hinks}},
  \bibinfo{author}{\bibfnamefont{D.~W.} \bibnamefont{Capone}},
  \bibinfo{author}{\bibfnamefont{K.}~\bibnamefont{Zhang}},
  \bibinfo{author}{\bibfnamefont{M.~B.} \bibnamefont{Brodsky}},
  \bibnamefont{and} \bibinfo{author}{\bibfnamefont{D.~J.}
  \bibnamefont{Scalapino}}, \bibinfo{journal}{Physical Review Letters}
  \textbf{\bibinfo{volume}{58}}, \bibinfo{pages}{1024} (\bibinfo{year}{1987}).

\bibitem[{\citenamefont{Lin et~al.}(2006)\citenamefont{Lin, Sahrakorpi,
  Markiewicz, and Bansil}}]{Lin2006}
\bibinfo{author}{\bibfnamefont{H.}~\bibnamefont{Lin}},
  \bibinfo{author}{\bibfnamefont{S.}~\bibnamefont{Sahrakorpi}},
  \bibinfo{author}{\bibfnamefont{R.~S.} \bibnamefont{Markiewicz}},
  \bibnamefont{and} \bibinfo{author}{\bibfnamefont{A.}~\bibnamefont{Bansil}},
  \bibinfo{journal}{Physical Review Letters} \textbf{\bibinfo{volume}{96}},
  \bibinfo{pages}{097001} (\bibinfo{year}{2006}).

\bibitem[{\citenamefont{Sahrakorpi et~al.}(2005)\citenamefont{Sahrakorpi,
  Lindroos, Markiewicz, and Bansil}}]{Sahrakorpi2005}
\bibinfo{author}{\bibfnamefont{S.}~\bibnamefont{Sahrakorpi}},
  \bibinfo{author}{\bibfnamefont{M.}~\bibnamefont{Lindroos}},
  \bibinfo{author}{\bibfnamefont{R.~S.} \bibnamefont{Markiewicz}},
  \bibnamefont{and} \bibinfo{author}{\bibfnamefont{A.}~\bibnamefont{Bansil}},
  \bibinfo{journal}{Physical Review Letters} \textbf{\bibinfo{volume}{95}},
  \bibinfo{pages}{157601} (\bibinfo{year}{2005}).

\bibitem[{\citenamefont{Yoshida et~al.}(2006)\citenamefont{Yoshida, Zhou,
  Tanaka, Yang, Hussain, Shen, Fujimori, Sahrakorpi, Lindroos, Markiewicz
  et~al.}}]{Yoshida2006}
\bibinfo{author}{\bibfnamefont{T.}~\bibnamefont{Yoshida}},
  \bibinfo{author}{\bibfnamefont{X.~J.} \bibnamefont{Zhou}},
  \bibinfo{author}{\bibfnamefont{K.}~\bibnamefont{Tanaka}},
  \bibinfo{author}{\bibfnamefont{W.~L.} \bibnamefont{Yang}},
  \bibinfo{author}{\bibfnamefont{Z.}~\bibnamefont{Hussain}},
  \bibinfo{author}{\bibfnamefont{Z.~X.} \bibnamefont{Shen}},
  \bibinfo{author}{\bibfnamefont{A.}~\bibnamefont{Fujimori}},
  \bibinfo{author}{\bibfnamefont{S.}~\bibnamefont{Sahrakorpi}},
  \bibinfo{author}{\bibfnamefont{M.}~\bibnamefont{Lindroos}},
  \bibinfo{author}{\bibfnamefont{R.~S.} \bibnamefont{Markiewicz}},
  \bibinfo{author}{\bibfnamefont{A.}~\bibnamefont{Bansil}},
  \bibinfo{author}{\bibfnamefont{S.}~\bibnamefont{Komiya}},
  \bibinfo{author}{\bibfnamefont{Y.}~\bibnamefont{Ando}},
  \bibinfo{author}{\bibfnamefont{H.}~\bibnamefont{Eisaki}},
  \bibinfo{author}{\bibfnamefont{T.}~\bibnamefont{Kakeshita}},
  \bibnamefont{and} \bibinfo{author}{\bibfnamefont{S.}~\bibnamefont{Uchida}},
  \bibinfo{journal}{Physical Review B} \textbf{\bibinfo{volume}{74}},
  \bibinfo{pages}{224510} (\bibinfo{year}{2006}).

\bibitem[{\citenamefont{Lanzara et~al.}(2001)\citenamefont{Lanzara, Bogdanov,
  Zhou, Kellar, Feng, Lu, Yoshida, Eisaki, Fujimori, Kishio
  et~al.}}]{Lanzara2001}
\bibinfo{author}{\bibfnamefont{A.}~\bibnamefont{Lanzara}},
  \bibinfo{author}{\bibfnamefont{P.~V.} \bibnamefont{Bogdanov}},
  \bibinfo{author}{\bibfnamefont{X.~J.} \bibnamefont{Zhou}},
  \bibinfo{author}{\bibfnamefont{S.~A.} \bibnamefont{Kellar}},
  \bibinfo{author}{\bibfnamefont{D.~L.} \bibnamefont{Feng}},
  \bibinfo{author}{\bibfnamefont{E.~D.} \bibnamefont{Lu}},
  \bibinfo{author}{\bibfnamefont{T.}~\bibnamefont{Yoshida}},
  \bibinfo{author}{\bibfnamefont{H.}~\bibnamefont{Eisaki}},
  \bibinfo{author}{\bibfnamefont{A.}~\bibnamefont{Fujimori}},
  \bibinfo{author}{\bibfnamefont{K.}~\bibnamefont{Kishio}},
  \bibinfo{author}{\bibfnamefont{J.~I.} \bibnamefont{Shimoyama}},
  \bibinfo{author}{\bibfnamefont{T.}~\bibnamefont{Noda}},
  \bibinfo{author}{\bibfnamefont{S.}~\bibnamefont{Uchida}},
  \bibinfo{author}{\bibfnamefont{Z.}~\bibnamefont{Hussain}}, \bibnamefont{and}
  \bibinfo{author}{\bibfnamefont{Z.~X.} \bibnamefont{Shen}},
  \bibinfo{journal}{Nature} \textbf{\bibinfo{volume}{412}},
  \bibinfo{pages}{510} (\bibinfo{year}{2001}).

\bibitem[{\citenamefont{Ronning et~al.}(2005)\citenamefont{Ronning, Shen,
  Armitage, Damascelli, Lu, Shen, Miller, and Kim}}]{Ronning2005}
\bibinfo{author}{\bibfnamefont{F.}~\bibnamefont{Ronning}},
  \bibinfo{author}{\bibfnamefont{K.~M.} \bibnamefont{Shen}},
  \bibinfo{author}{\bibfnamefont{N.~P.} \bibnamefont{Armitage}},
  \bibinfo{author}{\bibfnamefont{A.}~\bibnamefont{Damascelli}},
  \bibinfo{author}{\bibfnamefont{D.~H.} \bibnamefont{Lu}},
  \bibinfo{author}{\bibfnamefont{Z.~X.} \bibnamefont{Shen}},
  \bibinfo{author}{\bibfnamefont{L.~L.} \bibnamefont{Miller}},
  \bibnamefont{and} \bibinfo{author}{\bibfnamefont{C.}~\bibnamefont{Kim}},
  \bibinfo{journal}{Physical Review B} \textbf{\bibinfo{volume}{71}},
  \bibinfo{pages}{094518} (\bibinfo{year}{2005}).

\bibitem[{\citenamefont{Graf et~al.}(2007)\citenamefont{Graf, Gweon, McElroy,
  Zhou, Jozwiak, Rotenberg, Bill, Sasagawa, Eisaki, Uchida et~al.}}]{Graf2007}
\bibinfo{author}{\bibfnamefont{J.}~\bibnamefont{Graf}},
  \bibinfo{author}{\bibfnamefont{G.~H.} \bibnamefont{Gweon}},
  \bibinfo{author}{\bibfnamefont{K.}~\bibnamefont{McElroy}},
  \bibinfo{author}{\bibfnamefont{S.~Y.} \bibnamefont{Zhou}},
  \bibinfo{author}{\bibfnamefont{C.}~\bibnamefont{Jozwiak}},
  \bibinfo{author}{\bibfnamefont{E.}~\bibnamefont{Rotenberg}},
  \bibinfo{author}{\bibfnamefont{A.}~\bibnamefont{Bill}},
  \bibinfo{author}{\bibfnamefont{T.}~\bibnamefont{Sasagawa}},
  \bibinfo{author}{\bibfnamefont{H.}~\bibnamefont{Eisaki}},
  \bibinfo{author}{\bibfnamefont{S.}~\bibnamefont{Uchida}},
  \bibinfo{author}{\bibfnamefont{H.}~\bibnamefont{Takagi}},
  \bibinfo{author}{\bibfnamefont{D.~H.} \bibnamefont{Lee}}, \bibnamefont{and}
  \bibinfo{author}{\bibfnamefont{A.}~\bibnamefont{Lanzara}},
  \bibinfo{journal}{Physical Review Letters} \textbf{\bibinfo{volume}{98}},
  \bibinfo{pages}{067004} (\bibinfo{year}{2007}).

\bibitem[{\citenamefont{Meevasana et~al.}(2007)\citenamefont{Meevasana, Zhou,
  Sahrakorpi, Lee, Yang, Tanaka, Mannella, Yoshida, Lu, Chen
  et~al.}}]{Meevasana2007}
\bibinfo{author}{\bibfnamefont{W.}~\bibnamefont{Meevasana}},
  \bibinfo{author}{\bibfnamefont{X.~J.} \bibnamefont{Zhou}},
  \bibinfo{author}{\bibfnamefont{S.}~\bibnamefont{Sahrakorpi}},
  \bibinfo{author}{\bibfnamefont{W.~S.} \bibnamefont{Lee}},
  \bibinfo{author}{\bibfnamefont{W.~L.} \bibnamefont{Yang}},
  \bibinfo{author}{\bibfnamefont{K.}~\bibnamefont{Tanaka}},
  \bibinfo{author}{\bibfnamefont{N.}~\bibnamefont{Mannella}},
  \bibinfo{author}{\bibfnamefont{T.}~\bibnamefont{Yoshida}},
  \bibinfo{author}{\bibfnamefont{D.~H.} \bibnamefont{Lu}},
  \bibinfo{author}{\bibfnamefont{Y.~L.} \bibnamefont{Chen}},
  \bibinfo{author}{\bibfnamefont{R.~H.} \bibnamefont{He}},
  \bibinfo{author}{\bibfnamefont{H.}~\bibnamefont{Lin}},
  \bibinfo{author}{\bibfnamefont{S.}~\bibnamefont{Komiya}},
  \bibinfo{author}{\bibfnamefont{Y.}~\bibnamefont{Ando}},
  \bibinfo{author}{\bibfnamefont{F.}~\bibnamefont{Zhou}},
  \bibinfo{author}{\bibfnamefont{W.~X.} \bibnamefont{Ti}},
  \bibinfo{author}{\bibfnamefont{J.~W.} \bibnamefont{Xiong}},
  \bibinfo{author}{\bibfnamefont{Z.~X.} \bibnamefont{Zhao}},
  \bibinfo{author}{\bibfnamefont{T.}~\bibnamefont{Sasagawa}},
  \bibinfo{author}{\bibfnamefont{T.}~\bibnamefont{Kakeshita}},
  \bibinfo{author}{\bibfnamefont{K.}~\bibnamefont{Fujita}},
  \bibinfo{author}{\bibfnamefont{S.}~\bibnamefont{Uchida}},
  \bibinfo{author}{\bibfnamefont{H.}~\bibnamefont{Eisaki}},
  \bibinfo{author}{\bibfnamefont{A.}~\bibnamefont{Fujimori}},
  \bibinfo{author}{\bibfnamefont{Z.}~\bibnamefont{Hussain}},
  \bibinfo{author}{\bibfnamefont{R.~S.} \bibnamefont{Markiewicz}},
  \bibinfo{author}{\bibfnamefont{A.}~\bibnamefont{Bansil}},
  \bibinfo{author}{\bibfnamefont{N.}~\bibnamefont{Nagaosa}},
  \bibinfo{author}{\bibfnamefont{J.}~\bibnamefont{Zaanen}},
  \bibinfo{author}{\bibfnamefont{T.~P.} \bibnamefont{Devereaux}},
  \bibnamefont{and} \bibinfo{author}{\bibfnamefont{Z.~X.} \bibnamefont{Shen}},
  \bibinfo{journal}{Physical Review B} \textbf{\bibinfo{volume}{75}},
  \bibinfo{pages}{174506} (\bibinfo{year}{2007}).

\bibitem[{\citenamefont{Xie et~al.}(2007)\citenamefont{Xie, Yang, Shen, Zhao,
  Ou, Wei, Gu, Arita, Qiao, Namatame et~al.}}]{Xie2007}
\bibinfo{author}{\bibfnamefont{B.~P.} \bibnamefont{Xie}},
  \bibinfo{author}{\bibfnamefont{K.}~\bibnamefont{Yang}},
  \bibinfo{author}{\bibfnamefont{D.~W.} \bibnamefont{Shen}},
  \bibinfo{author}{\bibfnamefont{J.~F.} \bibnamefont{Zhao}},
  \bibinfo{author}{\bibfnamefont{H.~W.} \bibnamefont{Ou}},
  \bibinfo{author}{\bibfnamefont{J.}~\bibnamefont{Wei}},
  \bibinfo{author}{\bibfnamefont{S.~Y.} \bibnamefont{Gu}},
  \bibinfo{author}{\bibfnamefont{M.}~\bibnamefont{Arita}},
  \bibinfo{author}{\bibfnamefont{S.}~\bibnamefont{Qiao}},
  \bibinfo{author}{\bibfnamefont{H.}~\bibnamefont{Namatame}},
  \bibinfo{author}{\bibfnamefont{M.}~\bibnamefont{Taniguchi}},
  \bibinfo{author}{\bibfnamefont{N.}~\bibnamefont{Kaneko}},
  \bibinfo{author}{\bibfnamefont{H.}~\bibnamefont{Eisaki}},
  \bibinfo{author}{\bibfnamefont{K.~D.} \bibnamefont{Tsuei}},
  \bibinfo{author}{\bibfnamefont{C.~M.} \bibnamefont{Cheng}},
  \bibinfo{author}{\bibfnamefont{I.}~\bibnamefont{Vobornik}},
  \bibinfo{author}{\bibfnamefont{J.}~\bibnamefont{Fujii}},
  \bibinfo{author}{\bibfnamefont{G.}~\bibnamefont{Rossi}},
  \bibinfo{author}{\bibfnamefont{Z.~Q.} \bibnamefont{Yang}}, \bibnamefont{and}
  \bibinfo{author}{\bibfnamefont{D.~L.} \bibnamefont{Feng}},
  \bibinfo{journal}{Physical Review Letters} \textbf{\bibinfo{volume}{98}},
  \bibinfo{pages}{147001} (\bibinfo{year}{2007}).

\bibitem[{\citenamefont{Valla et~al.}(2007)\citenamefont{Valla, Kidd, Yin, Gu,
  Johnson, Pan, and Fedorov}}]{Valla2007}
\bibinfo{author}{\bibfnamefont{T.}~\bibnamefont{Valla}},
  \bibinfo{author}{\bibfnamefont{T.~E.} \bibnamefont{Kidd}},
  \bibinfo{author}{\bibfnamefont{W.~G.} \bibnamefont{Yin}},
  \bibinfo{author}{\bibfnamefont{G.~D.} \bibnamefont{Gu}},
  \bibinfo{author}{\bibfnamefont{P.~D.} \bibnamefont{Johnson}},
  \bibinfo{author}{\bibfnamefont{Z.~H.} \bibnamefont{Pan}}, \bibnamefont{and}
  \bibinfo{author}{\bibfnamefont{A.~V.} \bibnamefont{Fedorov}},
  \bibinfo{journal}{Physical Review Letters} \textbf{\bibinfo{volume}{98}},
  \bibinfo{pages}{167003} (\bibinfo{year}{2007}).

\bibitem[{\citenamefont{Sensarma et~al.}(2007)\citenamefont{Sensarma, Randeria,
  and Trivedi}}]{Sensarma2007}
\bibinfo{author}{\bibfnamefont{R.}~\bibnamefont{Sensarma}},
  \bibinfo{author}{\bibfnamefont{M.}~\bibnamefont{Randeria}}, \bibnamefont{and}
  \bibinfo{author}{\bibfnamefont{N.}~\bibnamefont{Trivedi}},
  \bibinfo{journal}{Physical Review Letters} \textbf{\bibinfo{volume}{98}},
  \bibinfo{pages}{027004} (\bibinfo{year}{2007}).

\bibitem[{\citenamefont{Kanigel et~al.}(2006)\citenamefont{Kanigel, Norman,
  Randeria, Chatterjee, Souma, Kaminski, Fretwell, Rosenkranz, Shi, Sato
  et~al.}}]{Kanigel2006}
\bibinfo{author}{\bibfnamefont{A.}~\bibnamefont{Kanigel}},
  \bibinfo{author}{\bibfnamefont{M.~R.} \bibnamefont{Norman}},
  \bibinfo{author}{\bibfnamefont{M.}~\bibnamefont{Randeria}},
  \bibinfo{author}{\bibfnamefont{U.}~\bibnamefont{Chatterjee}},
  \bibinfo{author}{\bibfnamefont{S.}~\bibnamefont{Souma}},
  \bibinfo{author}{\bibfnamefont{A.}~\bibnamefont{Kaminski}},
  \bibinfo{author}{\bibfnamefont{H.~M.} \bibnamefont{Fretwell}},
  \bibinfo{author}{\bibfnamefont{S.}~\bibnamefont{Rosenkranz}},
  \bibinfo{author}{\bibfnamefont{M.}~\bibnamefont{Shi}},
  \bibinfo{author}{\bibfnamefont{T.}~\bibnamefont{Sato}},
  \bibinfo{author}{\bibfnamefont{T.}~\bibnamefont{Takahashi}},
  \bibinfo{author}{\bibfnamefont{Z.~Z.} \bibnamefont{Li}},
  \bibinfo{author}{\bibfnamefont{H.}~\bibnamefont{Raffy}},
  \bibinfo{author}{\bibfnamefont{K.}~\bibnamefont{Kadowaki}},
  \bibinfo{author}{\bibfnamefont{D.}~\bibnamefont{Hinks}},
  \bibinfo{author}{\bibfnamefont{L.}~\bibnamefont{Ozyuzer}}, \bibnamefont{and}
  \bibinfo{author}{\bibfnamefont{J.~C.} \bibnamefont{Campuzano}},
  \bibinfo{journal}{Nature Physics} \textbf{\bibinfo{volume}{2}},
  \bibinfo{pages}{447} (\bibinfo{year}{2006}).

\bibitem[{\citenamefont{Bansil and Lindroos}(1999)}]{Bansil1999matrix}
\bibinfo{author}{\bibfnamefont{A.}~\bibnamefont{Bansil}} \bibnamefont{and}
  \bibinfo{author}{\bibfnamefont{M.}~\bibnamefont{Lindroos}},
  \bibinfo{journal}{Physical Review Letters} \textbf{\bibinfo{volume}{83}},
  \bibinfo{pages}{5154} (\bibinfo{year}{1999}).

\bibitem[{\citenamefont{Bansil et~al.}(2005)\citenamefont{Bansil, Lindroos,
  Sahrakorpi, and Markiewicz}}]{Bansil2005}
\bibinfo{author}{\bibfnamefont{A.}~\bibnamefont{Bansil}},
  \bibinfo{author}{\bibfnamefont{M.}~\bibnamefont{Lindroos}},
  \bibinfo{author}{\bibfnamefont{S.}~\bibnamefont{Sahrakorpi}},
  \bibnamefont{and} \bibinfo{author}{\bibfnamefont{R.~S.}
  \bibnamefont{Markiewicz}}, \bibinfo{journal}{Physical Review B}
  \textbf{\bibinfo{volume}{71}}, \bibinfo{pages}{012503}
  (\bibinfo{year}{2005}).

\bibitem{footnote}
See EPAPS Document No. E-PRBMDO-78-025834 for movies showing ARPES intensity maps at a series of energies passing through the VHS, at three dopings x=0.03, 0.07, and 0.12. These movies can be directly accessed at  \href{ftp://ftp.aip.org/epaps/phys_rev_b/E-PRBMDO-78-025834/}{ftp://ftp.aip.org/epaps/phys_rev_b/E-PRBMDO-78-025834/}. For more information on EPAPS, see \url{http://www.aip.org/pubservs/epaps.html}.

\bibitem[{\citenamefont{Zhou et~al.}(2003)\citenamefont{Zhou, Yoshida, Lanzara,
  Bogdanov, Kellar, Shen, Yang, Ronning, Sasagawa, Kakeshita
  et~al.}}]{Zhou2003}
\bibinfo{author}{\bibfnamefont{X.~J.} \bibnamefont{Zhou}},
  \bibinfo{author}{\bibfnamefont{T.}~\bibnamefont{Yoshida}},
  \bibinfo{author}{\bibfnamefont{A.}~\bibnamefont{Lanzara}},
  \bibinfo{author}{\bibfnamefont{P.~V.} \bibnamefont{Bogdanov}},
  \bibinfo{author}{\bibfnamefont{S.~A.} \bibnamefont{Kellar}},
  \bibinfo{author}{\bibfnamefont{K.~M.} \bibnamefont{Shen}},
  \bibinfo{author}{\bibfnamefont{W.~L.} \bibnamefont{Yang}},
  \bibinfo{author}{\bibfnamefont{F.}~\bibnamefont{Ronning}},
  \bibinfo{author}{\bibfnamefont{T.}~\bibnamefont{Sasagawa}},
  \bibinfo{author}{\bibfnamefont{T.}~\bibnamefont{Kakeshita}},
  \bibinfo{author}{\bibfnamefont{T.}~\bibnamefont{Noda}},
  \bibinfo{author}{\bibfnamefont{H.}~\bibnamefont{Eisaki}},
  \bibinfo{author}{\bibfnamefont{S.}~\bibnamefont{Uchida}},
  \bibinfo{author}{\bibfnamefont{C.~T.} \bibnamefont{Lin}},
  \bibinfo{author}{\bibfnamefont{F.}~\bibnamefont{Zhou}},
  \bibinfo{author}{\bibfnamefont{J.~W.} \bibnamefont{Xiong}},
  \bibinfo{author}{\bibfnamefont{W.~X.} \bibnamefont{Ti}},
  \bibinfo{author}{\bibfnamefont{Z.~X.} \bibnamefont{Zhao}},
  \bibinfo{author}{\bibfnamefont{A.}~\bibnamefont{Fujimori}},
  \bibinfo{author}{\bibfnamefont{Z.}~\bibnamefont{Hussain}}, \bibnamefont{and}
  \bibinfo{author}{\bibfnamefont{Z.~X.} \bibnamefont{Shen}},
  \bibinfo{journal}{Nature} \textbf{\bibinfo{volume}{423}},
  \bibinfo{pages}{398} (\bibinfo{year}{2003}).

\bibitem[{\citenamefont{Yoshida et~al.}(2003)\citenamefont{Yoshida, Zhou,
  Sasagawa, Yang, Bogdanov, Lanzara, Hussain, Mizokawa, Fujimori, Eisaki
  et~al.}}]{Yoshida2003}
\bibinfo{author}{\bibfnamefont{T.}~\bibnamefont{Yoshida}},
  \bibinfo{author}{\bibfnamefont{X.~J.} \bibnamefont{Zhou}},
  \bibinfo{author}{\bibfnamefont{T.}~\bibnamefont{Sasagawa}},
  \bibinfo{author}{\bibfnamefont{W.~L.} \bibnamefont{Yang}},
  \bibinfo{author}{\bibfnamefont{P.~V.} \bibnamefont{Bogdanov}},
  \bibinfo{author}{\bibfnamefont{A.}~\bibnamefont{Lanzara}},
  \bibinfo{author}{\bibfnamefont{Z.}~\bibnamefont{Hussain}},
  \bibinfo{author}{\bibfnamefont{T.}~\bibnamefont{Mizokawa}},
  \bibinfo{author}{\bibfnamefont{A.}~\bibnamefont{Fujimori}},
  \bibinfo{author}{\bibfnamefont{H.}~\bibnamefont{Eisaki}},
  \bibinfo{author}{\bibfnamefont{Z.~X.} \bibnamefont{Shen}},
  \bibinfo{author}{\bibfnamefont{T.}~\bibnamefont{Kakeshita}},
  \bibnamefont{and} \bibinfo{author}{\bibfnamefont{S.}~\bibnamefont{Uchida}},
  \bibinfo{journal}{Physical Review Letters} \textbf{\bibinfo{volume}{91}},
  \bibinfo{pages}{027001} (\bibinfo{year}{2003}).

\bibitem[{\citenamefont{Shen et~al.}(2004)\citenamefont{Shen, Ronning, Lu, Lee,
  Ingle, Meevasana, Baumberger, Damascelli, Armitage, Miller
  et~al.}}]{Shen2004}
\bibinfo{author}{\bibfnamefont{K.~M.} \bibnamefont{Shen}},
  \bibinfo{author}{\bibfnamefont{F.}~\bibnamefont{Ronning}},
  \bibinfo{author}{\bibfnamefont{D.~H.} \bibnamefont{Lu}},
  \bibinfo{author}{\bibfnamefont{W.~S.} \bibnamefont{Lee}},
  \bibinfo{author}{\bibfnamefont{N.~J.~C.} \bibnamefont{Ingle}},
  \bibinfo{author}{\bibfnamefont{W.}~\bibnamefont{Meevasana}},
  \bibinfo{author}{\bibfnamefont{F.}~\bibnamefont{Baumberger}},
  \bibinfo{author}{\bibfnamefont{A.}~\bibnamefont{Damascelli}},
  \bibinfo{author}{\bibfnamefont{N.~P.} \bibnamefont{Armitage}},
  \bibinfo{author}{\bibfnamefont{L.~L.} \bibnamefont{Miller}},
  \bibinfo{author}{\bibfnamefont{Y.}~\bibnamefont{Kohsaka}},
  \bibinfo{author}{\bibfnamefont{M.}~\bibnamefont{Azuma}},
  \bibinfo{author}{\bibfnamefont{M.}~\bibnamefont{Takano}},
  \bibinfo{author}{\bibfnamefont{H.}~\bibnamefont{Takagi}}, \bibnamefont{and}
  \bibinfo{author}{\bibfnamefont{Z.~X.} \bibnamefont{Shen}},
  \bibinfo{journal}{Physical Review Letters} \textbf{\bibinfo{volume}{93}},
  \bibinfo{pages}{267002} (\bibinfo{year}{2004}).

\bibitem[{\citenamefont{Rosch et~al.}(2005)\citenamefont{Rosch, Gunnarsson,
  Zhou, Yoshida, Sasagawa, Fujimori, Hussain, Shen, and Uchida}}]{Rosch2005}
\bibinfo{author}{\bibfnamefont{O.}~\bibnamefont{Rosch}},
  \bibinfo{author}{\bibfnamefont{O.}~\bibnamefont{Gunnarsson}},
  \bibinfo{author}{\bibfnamefont{X.~J.} \bibnamefont{Zhou}},
  \bibinfo{author}{\bibfnamefont{T.}~\bibnamefont{Yoshida}},
  \bibinfo{author}{\bibfnamefont{T.}~\bibnamefont{Sasagawa}},
  \bibinfo{author}{\bibfnamefont{A.}~\bibnamefont{Fujimori}},
  \bibinfo{author}{\bibfnamefont{Z.}~\bibnamefont{Hussain}},
  \bibinfo{author}{\bibfnamefont{Z.~X.} \bibnamefont{Shen}}, \bibnamefont{and}
  \bibinfo{author}{\bibfnamefont{S.}~\bibnamefont{Uchida}},
  \bibinfo{journal}{Physical Review Letters} \textbf{\bibinfo{volume}{95}},
  \bibinfo{pages}{227002} (\bibinfo{year}{2005}).

\bibitem[{\citenamefont{Paramekanti et~al.}(2004)\citenamefont{Paramekanti,
  Randeria, and Trivedi}}]{Paramekanti2004}
\bibinfo{author}{\bibfnamefont{A.}~\bibnamefont{Paramekanti}},
  \bibinfo{author}{\bibfnamefont{M.}~\bibnamefont{Randeria}}, \bibnamefont{and}
  \bibinfo{author}{\bibfnamefont{N.}~\bibnamefont{Trivedi}},
  \bibinfo{journal}{Physical Review B} \textbf{\bibinfo{volume}{70}},
  \bibinfo{pages}{054504} (\bibinfo{year}{2004}).

\bibitem[{\citenamefont{Matsuzaki et~al.}(2004)\citenamefont{Matsuzaki, Momono,
  Oda, and Ido}}]{Matsuzaki2004}
\bibinfo{author}{\bibfnamefont{T.}~\bibnamefont{Matsuzaki}},
  \bibinfo{author}{\bibfnamefont{N.}~\bibnamefont{Momono}},
  \bibinfo{author}{\bibfnamefont{M.}~\bibnamefont{Oda}}, \bibnamefont{and}
  \bibinfo{author}{\bibfnamefont{M.}~\bibnamefont{Ido}},
  \bibinfo{journal}{Journal of the Physical Society of Japan}
  \textbf{\bibinfo{volume}{73}}, \bibinfo{pages}{2232} (\bibinfo{year}{2004}).

\bibitem[{\citenamefont{Prelovsek and Ramsak}(2002)}]{Prelovsek2002}
\bibinfo{author}{\bibfnamefont{P.}~\bibnamefont{Prelovsek}} \bibnamefont{and}
  \bibinfo{author}{\bibfnamefont{A.}~\bibnamefont{Ramsak}},
  \bibinfo{journal}{Physical Review B} \textbf{\bibinfo{volume}{65}},
  \bibinfo{pages}{174529} (\bibinfo{year}{2002}).

\bibitem[{\citenamefont{Armitage et~al.}(2001)\citenamefont{Armitage, Lu, Kim,
  Damascelli, Shen, Ronning, Feng, Bogdanov, Shen, Onose
  et~al.}}]{Armitage2001}
\bibinfo{author}{\bibfnamefont{N.~P.} \bibnamefont{Armitage}},
  \bibinfo{author}{\bibfnamefont{D.~H.} \bibnamefont{Lu}},
  \bibinfo{author}{\bibfnamefont{C.}~\bibnamefont{Kim}},
  \bibinfo{author}{\bibfnamefont{A.}~\bibnamefont{Damascelli}},
  \bibinfo{author}{\bibfnamefont{K.~M.} \bibnamefont{Shen}},
  \bibinfo{author}{\bibfnamefont{F.}~\bibnamefont{Ronning}},
  \bibinfo{author}{\bibfnamefont{D.~L.} \bibnamefont{Feng}},
  \bibinfo{author}{\bibfnamefont{P.}~\bibnamefont{Bogdanov}},
  \bibinfo{author}{\bibfnamefont{Z.~X.} \bibnamefont{Shen}},
  \bibinfo{author}{\bibfnamefont{Y.}~\bibnamefont{Onose}},
  \bibinfo{author}{\bibfnamefont{Y.}~\bibnamefont{Taguchi}},
  \bibinfo{author}{\bibfnamefont{Y.}~\bibnamefont{Tokura}},
  \bibinfo{author}{\bibfnamefont{P.~K.} \bibnamefont{Mang}},
  \bibinfo{author}{\bibfnamefont{N.}~\bibnamefont{Kaneko}}, \bibnamefont{and}
  \bibinfo{author}{\bibfnamefont{M.}~\bibnamefont{Greven}},
  \bibinfo{journal}{Physical Review Letters} \textbf{\bibinfo{volume}{87}},
  \bibinfo{pages}{147003} (\bibinfo{year}{2001}).

\bibitem[{\citenamefont{Kusko et~al.}(2002)\citenamefont{Kusko, Markiewicz,
  Lindroos, and Bansil}}]{Kusko2002}
\bibinfo{author}{\bibfnamefont{C.}~\bibnamefont{Kusko}},
  \bibinfo{author}{\bibfnamefont{R.~S.} \bibnamefont{Markiewicz}},
  \bibinfo{author}{\bibfnamefont{M.}~\bibnamefont{Lindroos}}, \bibnamefont{and}
  \bibinfo{author}{\bibfnamefont{A.}~\bibnamefont{Bansil}},
  \bibinfo{journal}{Physical Review B} \textbf{\bibinfo{volume}{66}},
  \bibinfo{pages}{140513(R)} (\bibinfo{year}{2002}).

\bibitem[{\citenamefont{Tremblay et~al.}(2006)\citenamefont{Tremblay, Kyung,
  and Senechal}}]{Tremblay2006}
\bibinfo{author}{\bibfnamefont{A.~M.~S.} \bibnamefont{Tremblay}},
  \bibinfo{author}{\bibfnamefont{B.}~\bibnamefont{Kyung}}, \bibnamefont{and}
  \bibinfo{author}{\bibfnamefont{D.}~\bibnamefont{Senechal}},
  \bibinfo{journal}{Low Temperature Physics} \textbf{\bibinfo{volume}{32}},
  \bibinfo{pages}{424} (\bibinfo{year}{2006}).

\bibitem[{\citenamefont{Kokalj and Prelovsek}(2007)}]{Kokalj2007}
\bibinfo{author}{\bibfnamefont{J.}~\bibnamefont{Kokalj}} \bibnamefont{and}
  \bibinfo{author}{\bibfnamefont{P.}~\bibnamefont{Prelovsek}},
  \bibinfo{journal}{Physical Review B} \textbf{\bibinfo{volume}{75}},
  \bibinfo{pages}{045111} (\bibinfo{year}{2007}).

\bibitem[{\citenamefont{Edegger et~al.}(2006)\citenamefont{Edegger, Muthukumar,
  Gros, and Anderson}}]{Edegger2006}
\bibinfo{author}{\bibfnamefont{B.}~\bibnamefont{Edegger}},
  \bibinfo{author}{\bibfnamefont{V.~N.} \bibnamefont{Muthukumar}},
  \bibinfo{author}{\bibfnamefont{C.}~\bibnamefont{Gros}}, \bibnamefont{and}
  \bibinfo{author}{\bibfnamefont{P.~W.} \bibnamefont{Anderson}},
  \bibinfo{journal}{Physical Review Letters} \textbf{\bibinfo{volume}{96}},
  \bibinfo{pages}{207002} (\bibinfo{year}{2006}).

\bibitem[{\citenamefont{Ribeiro and Wen}(2005)}]{Ribeiro2005}
\bibinfo{author}{\bibfnamefont{T.~C.} \bibnamefont{Ribeiro}} \bibnamefont{and}
  \bibinfo{author}{\bibfnamefont{X.~G.} \bibnamefont{Wen}},
  \bibinfo{journal}{Physical Review Letters} \textbf{\bibinfo{volume}{95}},
  \bibinfo{pages}{057001} (\bibinfo{year}{2005}).

\bibitem[{\citenamefont{Tan et~al.}(2007)\citenamefont{Tan, Wan, and
  Wang}}]{Tan2007}
\bibinfo{author}{\bibfnamefont{F.}~\bibnamefont{Tan}},
  \bibinfo{author}{\bibfnamefont{Y.}~\bibnamefont{Wan}}, \bibnamefont{and}
  \bibinfo{author}{\bibfnamefont{Q.~H.} \bibnamefont{Wang}},
  \bibinfo{journal}{Physical Review B} \textbf{\bibinfo{volume}{76}},
  \bibinfo{pages}{054505} (\bibinfo{year}{2007}).

\end{thebibliography}
\end{document}